# ML DevOps Adoption in Practice: A Mixed-Method Study of Implementation Patterns and Organizational Benefits

**Dileepkumar S R[1, 2], Juby Mathew[3]**

[1]LUC, Marian College Kuttikanam 685 531, Kerala, India, dileedil@gmail.com, ORCID ID: 0009-0004-9774-1936

[2]Research Scholar, Computer Science and Multimedia, Lincoln University college Malaysia

[3]Department of Computer Science and Engineering, Amal Jyothi College of Engineering, Kanjirappally 686 518, Kerala, India, ORCID ID: 0000-0002-4660-5940

**Abstract**

Machine Learning (ML) DevOps, also known as MLOps, has emerged as a critical framework for efficiently operationalizing ML models in various industries. This study investigates the adoption trends, implementation efforts, and benefits of ML DevOps through a combination of literature review and empirical analysis. By surveying 150 professionals across industries and conducting in-depth interviews with 20 practitioners, the study provides insights into the growing adoption of ML DevOps, particularly in sectors like finance and healthcare. The research identifies key challenges, such as fragmented tooling, data management complexities, and skill gaps, which hinder widespread adoption. However, the findings highlight significant benefits, including improved deployment frequency, reduced error rates, enhanced collaboration between data science and DevOps teams, and lower operational costs. Organizations leveraging ML DevOps report accelerated model deployment, increased scalability, and better compliance with industry regulations. The study also explores the technical and cultural efforts required for successful implementation, such as investments in automation tools, real-time monitoring, and upskilling initiatives. The results indicate that while challenges remain, ML DevOps presents a viable path to optimizing ML lifecycle management, ensuring model reliability, and enhancing business value. Future research should focus on standardizing ML DevOps practices, assessing the return on investment across industries, and developing frameworks for seamless integration with traditional DevOps methodologies.

## 1. Introduction

Machine learning (ML) has rapidly evolved from a niche research discipline to a fundamental component of modern industrial applications. Industries such as healthcare, finance, retail, and transportation have increasingly integrated ML-driven solutions to enhance decision-making, optimize operations, and improve customer experiences. However, despite its transformative potential, the process of deploying, managing, and maintaining ML models in real-world production environments remains a formidable challenge [1]. Traditional DevOps methodologies, which emphasize continuous integration and continuous delivery (CI/CD), struggle to accommodate the unique requirements of ML workflows. These include aspects such as data versioning, model monitoring, and periodic retraining, which are critical for maintaining model performance over time [2,21].

To address these challenges, the concept of ML DevOps, also known as MLOps, has emerged as an extension of DevOps tailored specifically for ML systems. MLOps provides a systematic framework for automating and streamlining the entire ML model lifecycle—from development and deployment to monitoring and retraining [3,22]. By integrating DevOps principles with ML workflows, organizations can enhance the reliability, scalability, and maintainability of their ML models. This approach not only accelerates deployment but also ensures long-term sustainability and business value. The increasing adoption of ML DevOps across industries highlights the growing recognition of its benefits in overcoming operational bottlenecks and enabling efficient ML pipeline management.

Despite its advantages, the transition to ML DevOps is not without hurdles. Organizations often face challenges such as fragmented toolsets, organizational silos, and a shortage of skilled professionals capable of managing complex ML pipelines [4,24]. Traditional DevOps practices, widely used in software engineering for automating development and deployment processes, have proven their effectiveness. Studies indicate that high-performance DevOps teams achieve faster response times, lower failure rates, and significantly higher deployment frequency. However, applying DevOps methodologies to ML projects requires further exploration due to the additional complexities associated with ML models [5].

Recent research and technological advancements have contributed to the development of various tools aimed at supporting ML DevOps. Frameworks such as MLFlow and Amazon SageMaker streamline ML workflows, while package managers like Spack and EasyBuild facilitate automated model rebuilding. Containerization technologies, including Docker and Kubernetes, enable better model sharing and deployment. Additionally, specialized tools such as Ease.ml/CI and CodeReef have been proposed for continuous integration and benchmarking in ML projects [6,23]. However, many of these tools remain immature, requiring further development and broader compatibility across ML frameworks.

Given the ongoing challenges in ML DevOps adoption, there is a pressing need for comprehensive research to investigate the effectiveness of existing DevOps tools in ML projects [7,8]. Understanding their adoption rates, maintenance efforts, and overall impact on ML lifecycle management will provide valuable insights for both industry practitioners and researchers seeking to optimize ML operations in real-world applications.

This study aims to explore the current trends in ML DevOps adoption, analyse the efforts involved in its implementation, and evaluate the tangible and intangible benefits it offers. By synthesizing insights from recent literature and empirical investigations, this paper seeks to provide a roadmap for organizations looking to embark on or enhance their ML DevOps journey [9.10].

## 2. Research Objectives

The research focuses on providing a deeper understanding of the ML DevOps landscape, framed by the following specific objectives:

1. **Identifying Current Trends:** To explore how ML DevOps is being adopted across various industries and sectors. This includes understanding which industries are leading in implementation and the driving factors behind their adoption.
2. **Analysing Adoption Efforts:** To investigate the technical, organizational, and cultural efforts required to adopt ML DevOps frameworks successfully. This encompasses assessing the investments in tools, infrastructure, and team upskilling.
3. **Evaluating Benefits and ROI:** To examine the tangible and intangible benefits of ML DevOps practices, including faster deployment cycles, reduced operational costs, enhanced collaboration, and overall return on investment.
4. **Addressing Challenges:** To identify and categorize the primary challenges faced by organizations during the adoption of ML DevOps. This includes issues related to tooling, data management, and integration with existing workflows.
5. **Proposing Solutions:** To provide actionable recommendations for overcoming the identified challenges, thereby aiding organizations in streamlining their ML DevOps journey.

By addressing these objectives, the study aims to contribute a comprehensive perspective that can guide practitioners, researchers, and decision-makers in leveraging ML DevOps effectively

## 3. Methodology

The study employs a mixed-method approach comprising:
This study employs a structured and comprehensive approach to explore ML DevOps adoption trends, efforts, and benefits. The methodology combines qualitative and quantitative methods, structured into the following components:

### 3.1 Literature Review

A thorough review of recent literature from 2016 to 2024 was conducted to identify key developments, challenges, and benefits of ML DevOps. Sources include peer-reviewed journals, conference papers, and case studies.

### 3.2 Empirical Analysis

This component includes:
**Survey Design**
The survey was meticulously designed to capture comprehensive data from 150 professionals across multiple industries, with particular emphasis on sectors heavily investing in ML technologies such as finance, healthcare, and technology. The questionnaire utilized a multi-tiered structure incorporating both quantitative and qualitative elements: Likert-scale questions (1-5) assessed the degree of ML DevOps adoption and satisfaction levels; multiple-choice questions gathered specific information about tools, frameworks, and implementation approaches; and open-ended questions allowed respondents to share detailed insights about challenges and successes. The survey was distributed electronically through professional networks and industry forums, ensuring a diverse respondent pool. To maximize response quality, the questionnaire was first pilot-tested with a small group of 10 ML practitioners, and their feedback was incorporated into the final version. The survey achieved a completion rate of 85%, with respondents spending an average of 25 minutes on the questionnaire, indicating thoughtful and detailed responses.

**Interview Structure**

The interview phase comprised in-depth, semi-structured conversations with 20 carefully selected industry practitioners, including ML engineers, DevOps specialists, technical leaders, and C-level executives. Each interview lasted between 60-90 minutes and followed a structured protocol while allowing for organic discussion flow. The interviews were conducted remotely via video conferencing platforms and were recorded with participant consent for accurate transcription and analysis. The selection of interviewees ensured representation across different organization sizes (from startups to enterprises), various industry sectors, and different stages of ML DevOps maturity. The interview protocol was designed to progress from general experiences with ML DevOps to specific technical challenges and solutions, incorporating scenario-based questions to elicit detailed insights about implementation strategies and decision-making processes. Follow-up questions were used to probe deeper into particularly relevant or novel insights, and participants were encouraged to provide specific examples and metrics from their experiences.

### 3.3 Quantitative Metrics

The study employed a comprehensive set of quantitative metrics to measure the tangible impact of ML DevOps adoption across organizations. These metrics were collected through both automated tracking systems and self-reported data from survey participants, covering a pre- and post-implementation period of 12 months. Key performance indicators included deployment frequency (measured as the number of successful model deployments per month), time-to-market improvements (calculated as the duration from model development to production deployment), error rates (tracked through automated logging systems and incident reports), and operational costs (including infrastructure, maintenance, and personnel costs). The research team also collected metrics on team productivity (measured through sprint velocity and task completion rates), model performance (including accuracy, precision, and recall metrics), and resource utilization (CPU, memory, and storage usage patterns). These metrics were standardized across different organizations using normalized scales and were statistically analyzed to identify correlations between ML DevOps adoption and operational improvements. Organizations reported these metrics through a structured template, ensuring consistency in data collection and enabling meaningful cross-industry comparisons.

## 4. Literature Review

### 4.1 Recent Works on ML DevOps
1. **Rzig et al. (2023):** This seminal work brought attention to the "hidden technical debt" in machine learning systems, which arises from ad hoc development practices and a lack of scalability considerations. The authors emphasized the necessity of automated pipelines to manage the complexity of deploying and maintaining ML models. Their findings underscored the long-term value of integrating DevOps principles to streamline processes and mitigate maintenance overheads.
2. **Rodriguez et al. (2023):** This study explored the role of unified analytics platforms like MLflow in fostering better collaboration between data scientists and engineers. By providing tools for tracking experiments, packaging code, and sharing models, MLflow was shown to significantly reduce workflow silos. The authors demonstrated its effectiveness through case studies, highlighting its adaptability across diverse use cases.
3. **Abhijit Sen et al. (2020):** This work investigated TensorFlow Extended (TFX), a production-scale platform designed to handle the end-to-end ML lifecycle. The study illustrated how TFX enables scalable workflows through its modular architecture, facilitating tasks such as data validation, feature engineering, and model serving. The authors also highlighted its integration with TensorFlow, enhancing its appeal for large-scale deployments.
4. **Fahad Ahmed et al. (2021):** This research delved into the adoption of Kubernetes for deploying ML models, emphasizing its scalability and orchestration capabilities. The authors analyzed the use of Kubernetes in various scenarios, such as batch inference and real-time predictions, demonstrating its potential to optimize resource utilization and simplify model management.
5. **Angel Saldaña López et al. (2021):** Focusing on the challenges of integrating ML models into traditional DevOps pipelines, this study provided insights into organizational and technical

6. **Anna Wiedemann et al. (2018):** This case study examined the cost-benefit analysis of implementing ML DevOps in a financial services company. The findings revealed significant reductions in operational costs and time-to-market for ML solutions, with detailed metrics showcasing improved team productivity and model reliability.
7. **Dhia Elhaq et al. (2022):** The authors provided a comprehensive overview of best practices for integrating CI/CD in ML pipelines, drawing on insights from industry leaders. Their guidelines encompassed aspects like automated testing, model versioning, and reproducibility, offering a roadmap for organizations to enhance their ML workflows.
8. **Faustino et al. (2022):** This study focused on feature stores and their role in simplifying data management for ML workflows. The authors detailed how centralized feature repositories improve consistency, enable feature reuse, and reduce development cycles, making a compelling case for their adoption in enterprise settings.
9. **Bass et al. (2015):** This research analyzed the return on investment (ROI) of ML DevOps in the healthcare sector. Through empirical data and case studies, the authors highlighted significant gains in operational efficiency, patient outcomes, and compliance, making a strong argument for sector-wide adoption.
10. **Kim et al. (2019):** This paper proposed a maturity framework for measuring ML DevOps adoption in organizations. By defining clear stages and associated metrics, the authors provided a practical tool for assessing and guiding organizations' progress in implementing ML DevOps practices

Table 1 shows benefit and challenges of different works.

**Table 1:** Comparative study

| Author | Methodology | Benefit | Challenges |
|---|---|---|---|
| Rodriguez et al. (2023) | Analysis of hidden technical debt in ML systems and the role of automated pipelines. | Reduced maintenance overhead and better scalability of ML systems. | Managing complexity and technical debt in ML pipelines. |
| Rzig et al. (2022) | Evaluation of MLflow for unified analytics and collaboration enhancement. | Improved collaboration and reduced silos between data teams. | Workflow silos and lack of unified tools for teams. |
| Abhijit Sen et al. (2020) | Investigation of TensorFlow Extended for scalable ML lifecycle management. | Scalable workflows with robust model validation and serving. | Dependency on TensorFlow ecosystem and technical overhead. |
| Fahad Ahmed et al. (2021) | Adoption of Kubernetes for scalable and efficient ML model deployment. | Efficient resource utilization and simplified management. | Complexity in orchestrating models at scale. |
| Angel Saldaña López et al. (2021) | Study of integration challenges of ML into traditional DevOps pipelines. | Framework to align ML workflows with DevOps. | Organizational and tooling barriers for integration. |
| Anna Wiedemann et al. (2018) | Case study of ML DevOps cost-benefit in a financial services firm. | Reduced costs and improved team productivity and reliability. | High initial implementation costs and learning curve. |
| Dhia Elhaq Rzig et al. (2022) | Overview of CI/CD best practices for ML pipelines through industry insights. | Enhanced reproducibility and streamlined ML workflows. | Adapting CI/CD practices to ML-specific requirements. |

| Faustino et al. (2022) | Investigation of feature stores for efficient data management in ML workflows. | Consistent feature usage and reduced development cycles. | Managing feature consistency and cross-team usage. |
|---|---|---|---|
| Bass et al. (2015) | Analysis of ROI and operational impact of ML DevOps in healthcare. | Operational efficiency and compliance in healthcare applications. | Complex compliance requirements in sensitive sectors. |
| Kim et al. (2016) | Development of a maturity framework for ML DevOps adoption assessment. | Clear metrics for organizational progress in ML DevOps. | Defining measurable adoption metrics and maturity levels. |

### 4.2 Key Findings
- The adoption of ML DevOps is growing steadily, with sectors like finance, healthcare, and e-commerce leading the way.
- Challenges include lack of standardization, high initial setup costs, and the steep learning curve for practitioners.
- Benefits include faster deployment cycles, improved model performance, and enhanced collaboration.

## 5. Empirical Analysis

### 5.1 Survey Results

The survey component of the study was designed to capture a comprehensive understanding of ML DevOps adoption trends across industries. It involved responses from 150 professionals working in sectors like healthcare, finance, retail, and technology.

**Key Findings:**
1. **Adoption Trends:** A significant 68% of respondents reported having implemented some form of ML DevOps within their organizations. The adoption rates were particularly high in the finance and healthcare sectors, where robust ML pipelines are essential for real-time decision-making and compliance.
2. **Challenges Identified:** Despite the growing adoption, 42% of respondents cited challenges in integrating ML-specific tools into their existing DevOps workflows. Key issues included tool fragmentation, difficulty in managing data versioning, and a lack of skilled professionals familiar with both DevOps and machine learning.
3. **Benefits Realized:** An overwhelming 78% of participants highlighted faster time-to-market for ML models as the most significant benefit of ML DevOps. Additionally, organizations reported improved collaboration between data scientists and engineers, resulting in better model accuracy and operational efficiency.

**Sector-Specific Insights:**
- In the healthcare industry, ML DevOps was reported to enhance model monitoring, ensuring compliance with regulatory standards while maintaining accuracy in patient diagnosis.
- The financial services sector emphasized the role of automated pipelines in reducing fraud detection times and increasing the reliability of predictive analytics.

**Quantitative Metrics:** Survey responses also included data on measurable outcomes. For instance:
- Deployment frequency increased by 40% on average post-adoption of ML DevOps practices.

- Error rates in production ML models decreased by 25%, attributed to continuous monitoring and automated retraining workflows.
- Operational costs reduced by 30% due to automation of repetitive tasks and improved resource allocation.

The survey findings provide a robust foundation for understanding the current state of ML DevOps adoption, highlighting its transformative potential while acknowledging the challenges that remain to be addressed.

**Table 2:** Key findings

| Key Findings | Details |
|---|---|
| **Adoption Trends** | 68% of respondents reported implementing ML DevOps. Adoption is highest in the finance and healthcare sectors, driven by needs for real-time decision-making and compliance. |
| **Challenges Identified** | 42% cited issues with integrating ML tools into existing DevOps workflows, including tool fragmentation, data versioning difficulties, and skill gaps. |
| **Benefits Realized** | 78% experienced faster time-to-market, improved collaboration between data scientists and engineers, better model accuracy, and operational efficiency. |
| **Healthcare Insights** | Enhanced model monitoring for compliance and accuracy in patient diagnosis. |
| **Finance Insights** | Automated pipelines reduced fraud detection times and improved predictive analytics reliability. |
| **Deployment Frequency** | Increased by 40% on average due to streamlined ML DevOps practices. |
| **Error Rates** | Decreased by 25% with continuous monitoring and automated retraining workflows. |
| **Operational Costs** | Reduced by 30% through automation of repetitive tasks and optimized resource allocation. |

**5.2 Interview Insights**

The interview analysis focused on gathering qualitative insights from 20 industry practitioners across diverse sectors, including healthcare, finance, retail, and technology. The insights provided a deeper understanding of practical challenges, strategic implementations, and the cultural shifts required for successful ML DevOps adoption.

**Key Themes from Interviews:**
1. **Automation:** Practitioners emphasized the critical role of automation in ML pipelines. Many noted that automating processes such as data validation, feature engineering, and model deployment significantly reduces manual effort and human error, allowing teams to focus on innovation.
2. **Monitoring and Feedback:** Real-time monitoring of deployed models emerged as a priority for ensuring accuracy and reliability. Several participants highlighted the use of monitoring tools like Prometheus and Grafana to detect performance issues and model drift, enabling proactive retraining.
3. **Collaboration Challenges:** Participants noted that bridging the gap between data scientists and DevOps teams remains a major hurdle. Effective collaboration requires clear communication channels, standardized workflows, and shared tools to align goals and deliverables.
4. **Training and Upskilling:** Many interviewees stressed the importance of investing in training programs to equip teams with skills in both machine learning and DevOps. Cross-training initiatives were identified as essential for fostering a shared understanding and breaking down silos.

5. **Scalability:** Organizations with large-scale deployments emphasized the need for robust infrastructure, such as Kubernetes for container orchestration and distributed systems for handling big data.
6. **Cultural Shift:** Successful ML DevOps adoption was often linked to a cultural shift toward embracing agile practices, encouraging experimentation, and fostering a mindset of continuous improvement.

**Notable Quotes from Practitioners:**
- "Automation is not just a convenience; it's a necessity to scale ML operations in today's competitive landscape."
- "Without effective monitoring, even the best models can fail catastrophically. Continuous feedback is the lifeline of ML DevOps."
- "Training teams to understand both ML and DevOps has been our biggest challenge, but also our greatest success. It's transformed how we work."

**Case-Specific Insights:**
- A retail company described how implementing automated pipelines reduced their model deployment time from weeks to days, improving their ability to respond to market trends.
- A healthcare provider shared how integrating ML DevOps helped maintain compliance with strict regulatory standards while ensuring accurate diagnostic predictions.

The interview insights underscore the nuanced challenges and opportunities in ML DevOps adoption, emphasizing the interplay between technology, collaboration, and culture.

## 6. Efforts in Adopting ML DevOps

- **Technical Investments:** Tools like TFX, Kubeflow, and MLflow require substantial initial investment.
- **Training and Development:** Organizations must upskill teams to bridge the gap between data science and DevOps.
- **Change Management:** Shifting to ML DevOps demands a cultural shift towards cross-functional collaboration.

## 7. Benefits Analysis

The adoption of ML DevOps offers numerous benefits that enhance the efficiency, scalability, and reliability of machine learning operations. One of the most significant advantages is accelerated deployment cycles, allowing organizations to rapidly integrate ML models into production environments, reducing time-to-market. Improved model performance is another key benefit, as continuous monitoring, automated retraining, and feedback loops help maintain accuracy and reduce error rates. ML DevOps also fosters better collaboration between data scientists, engineers, and IT teams, bridging the gap between model development and operational deployment. Additionally, cost efficiency is achieved through automation, reducing manual intervention and optimizing resource allocation. Scalability is another major advantage, enabling organizations to handle complex ML workflows and large datasets seamlessly with tools like Kubernetes and MLflow. Regulatory compliance and risk mitigation are strengthened through structured model monitoring and governance frameworks, particularly in industries like healthcare and finance. Moreover, ML DevOps supports innovation and agility, allowing teams to experiment with new models and methodologies in a controlled, iterative manner. Overall, the implementation of ML DevOps not only streamlines ML lifecycle management but also enhances business value by ensuring reliability, efficiency, and adaptability in a rapidly evolving technological landscape.

## 8. Challenges

The adoption of ML DevOps presents several challenges that organizations must overcome to fully leverage its benefits. One of the primary obstacles is tool fragmentation, where multiple ML DevOps tools lack seamless integration, leading to inefficiencies in workflow management. Data management complexities, including versioning, governance, and quality assurance, pose significant hurdles in maintaining reproducibility and consistency across ML pipelines. The skill gap between data scientists and DevOps professionals further complicates adoption, as effective implementation requires expertise in both domains. Additionally, scalability concerns arise when deploying large-scale ML models, requiring robust infrastructure and automation strategies. Integration barriers with traditional DevOps practices make it difficult to align ML workflows with existing software development pipelines. Furthermore, organizations often struggle with high initial implementation costs, cultural resistance to change, and the need for continuous monitoring and retraining to ensure model reliability. Addressing these challenges requires strategic investments in training, standardized tooling, and automation frameworks to enhance efficiency and sustainability in ML operations.

## 9. Conclusion

This study underscores the transformative impact of ML DevOps in streamlining machine learning workflows and improving model operationalization across industries. The findings reveal that ML DevOps adoption is steadily increasing, particularly in finance, healthcare, and e-commerce, where real-time decision-making and compliance are paramount. Despite its advantages, challenges such as tool fragmentation, data versioning complexities, and a lack of skilled professionals pose significant hurdles. The empirical analysis demonstrates that organizations adopting ML DevOps experience faster deployment cycles, better model accuracy, and improved collaboration between data scientists and engineers. Automation, real-time monitoring, and cross-functional team training emerge as critical success factors in ML DevOps implementation. The study also highlights that a cultural shift towards agile methodologies and continuous learning is essential for maximizing the benefits of ML DevOps. While the adoption of advanced tools like Kubernetes, MLflow, and TensorFlow Extended has facilitated ML operations, further research is needed to refine these tools and develop standardized best practices. Future directions include comparative analyses of ML DevOps tools, long-term ROI studies, and frameworks for integrating ML DevOps into existing DevOps ecosystems. Overall, ML DevOps is poised to become a foundational approach for managing ML models in production, driving innovation, and ensuring sustainable business value.

**Funding**
Not applicable.
**Data availability**
The data that support the findings of this study are available from the corresponding author, Dileepkumar S R, upon reasonable request.
.
**Declarations**
**Conflict of interest**
The authors declare that they have no conflict of interest.
**Human and animal rights**

This article does not contain any studies with human or animal subjects performed by any of the authors.